\documentclass{article}

\begin{document}

\author{J.J. Rosales$ \footnote{E-mail:
rosales@salamanca.ugto.mx}$\\
Facultad de Ingenier\'{\i}a Mec\'anica, El\'ectrica y
Electr\'onica. \\
Campus FIMEE, Universidad de Gto.\\
Carretera Salamanca-Valle de Santiago, km. 3.5 + 1.8 km.\\
Comunidad de Palo Blanco, Salamanca Gto. M\'exico.\\
\\
V.I. Tkach$ \footnote{E-mail:
tkach@northwestern.edu}$\\
Department of Physics and Astronomy\\
Northwestern University\\
Evanston, IL 60208-3112, USA}
\title{Supersymmetric Cosmology and Dark Energy}
\maketitle

{\bf Abstract:} Using the superfield approach we construct the
$n=2$ supersymmetric lagrangian for the FRW Universe with perfect
fluid as matter fields. The obtained supersymmetric algebra
allowed us to take the square root of the Wheeler-DeWitt equation
and solve the corresponding quantum constraint. This model leads
to the relation between the vacuum energy density and the energy
density of the dust matter.
\\

\begin{center}
{\bf Introduction}
\end{center}
This paper is for the anniversary volume on the occasion 50th
birthday, Sergei Odintsov, our colleague and friend who made an
extensive contribution to the cosmological and astrophysics
fields.

Some time ago we have used the superfield formulation to
investigate supersymmetric cosmological models \cite{1}. The main
idea is to extend the group of local time reparametrization of the
cosmological models to the local time supersymmetry which is a
subgroup of the four dimensional space-time supersymmetry. This
local supersymmetry procedure has the advantage that, by defining
the superfields on superspace, all the component fields in a
supermultiplet can be manipulated simultaneously in a manner that
automatically preserves supersymmetry. Besides, the fermionic
fields are obtained in a clear manner as the supersymmetric
partners of the cosmological bosonic variables.
\\

More recently, using the superfield formulation the canonical
procedure quantization for a closed FRW cosmological model filled
with pressureless matter (dust) content and the corresponding
superpartner was reported \cite{2}. We have obtained the
quantization for the energy-like parameter, and it was shown, that
this energy is associated with the mass parameter quantization,
and that such type of Universe has a quantized masses of the order
of the Planck mass.
\\

In the present work we are interested in the construction of the
$n=2$ supersymmetric lagrangian for the FRW Universe with
barotropic perfect fluid as matter field including the
cosmological constant. The simplest dark energy candidate is the
cosmological constant stemming from energy density of the vacuum
\cite{3}. The obtained supersymmetric algebra allowed us to take
the square root of the Wheeler-DeWitt equation and solve the
corresponding quantum constraint.
\begin{center}
{\bf Classical Action}
\end{center}
The classical action for a pure gravity system and the
corresponding term of matter content, perfect fluid with a
constant equation of state parameter $\gamma$; $p = \gamma \rho$,
and the cosmological term is \cite{2}
\begin{equation}
S = \int\Big[ -\frac{c^2 R}{2N \tilde G}\Big(\frac{dR}{dt}\Big)^2
+ \frac{Nkc^4}{2\tilde G}R + \frac{Nc^4 \Lambda}{6 \tilde G} R^3 +
NM_{\gamma}c^2 R^{-3\gamma}\Big] dt. \label{2}
\end{equation}
where $c$ is the velocity of light in vacuum, $ \tilde G = \frac{8
\pi G}{6}$ where $G$ is the Newtonian gravitational constant; $k =
1,0, -1$ stands for spherical, plane or hyperspherical three
space; $N(t), R(t)$ are the lapse function and the scale factor,
respectively; $M_\gamma$ is the mass by unit
${\rm length}^{-\gamma}$.\\

The purpose of this work is the supersymmetrization of the full
action (\ref{2}) using the superfield approach. The action
(\ref{2}) is invariant under the time reparametrization
\begin{equation}
t^{\prime} \to t + a(t), \label{3}
\end{equation}
if the transformations of $R(t)$ and $N(t)$ are defined as
\begin{equation}
\delta R = a {\dot R}, \qquad \delta N = (aN)^{.}\label{4}
\end{equation}
The variation with respect to $R(t)$ and $N(t)$ lead to the
classical equation for the scale factor $R(t)$ and the constraint,
which generates the local reparametrization of $R(t)$ and $N(t)$.
This constraint leads to the Wheeler-DeWitt equation in quantum
cosmology.
\\

In order to obtain the corresponding supersymmetric action for
(\ref{2}), we follow the superfield approach. For this, we extend
the transformation of time reparametrization (\ref{3}) to the $n=
2$ local supersymmetry of time $(t, \eta, \bar\eta)$. Then, we
have the following local supersymmetric transformation
\begin{eqnarray} \delta{t}  &  = & a(t) +
\frac{i}{2}[\eta\beta^{\prime}(t) + \bar
\eta{\bar\beta^{\prime}(t)}],\nonumber\\
\delta\eta &  = & \frac{1}{2}\bar\beta^{\prime}(t)+ \frac{1}{2}
[\dot a(t) + ib(t)]\eta +
\frac{i}{2}\dot{\bar\beta}^{\prime}(t)\eta\bar\eta
,\label{5}\\
\delta{\bar\eta}  &  = & \frac{1}{2}\beta^{\prime}(t) +
\frac{1}{2}[\dot a(t) - ib(t)]\bar\eta- \frac{i}{2}
\dot\beta^{\prime}(t) \eta\bar\eta,\nonumber
\end{eqnarray}
where $\eta$ is a complex odd parameter ($\eta$ odd ``time''
coordinates), $\beta^{\prime}(t) = N^{-1/2}\beta(t)$ is the
Grassmann complex parameter of the local ``small'' $n=2$
supersymmetry (SUSY) transformation, and $b(t)$ is the parameter
of local $U(1)$ rotations of the complex $\eta$.
\\

For the closed $(k = 1)$ and plane $(k = 0)$ FRW action we propose
the following superfield generalization of the action (\ref{2}),
invariant under the $n = 2$ local supersymmetric transformation
(\ref{5})
\begin{eqnarray}
S_{susy} &=& \int \Big[ -\frac{c^2}{2 \tilde G} {I\!\!N}^{-1}
{I\!\!R}D_{\bar\eta}{I\!\!R} D_\eta{I\!\!R} + \frac{c^3 {\sqrt
k}}{2 \tilde G} {I\!\!R}^2 + \frac{c^3
\Lambda^{1/2}}{3\sqrt{3}\tilde G}{I\!\!R}^3 -\nonumber\\
&-& \frac{2 \sqrt{2}M^{1/2}_\gamma}{(3 - 3\gamma) \tilde G^{1/2}}
{I\!\!R}^{\frac{3-3\gamma}{2}}\Big]d{\eta} d{\bar\eta}dt,
\label{6}
\end{eqnarray}
where
\begin{equation}
D_{\eta} = \frac{\partial }{\partial \eta} + i\bar\eta
\frac{\partial }{\partial t}, \qquad D_{\bar\eta} =
-\frac{\partial }{\partial \bar\eta} - i\eta \frac{\partial
}{\partial t}, \label{7}
\end{equation}
are the supercovariant derivatives of the global "small"
supersymmetry of the generalized parameter corresponding to $t$.
The local supercovariant derivatives have the form ${\tilde
D}_{\eta} = {I\!\!N}^{-1/2} D_\eta$, ${\tilde D}_{\bar\eta} =
{I\!\!N}^{-1/2} D_{\bar\eta}$, and ${I\!\!R}(t,\eta, \bar\eta),
{I\!\!N}(t,\eta, \bar\eta)$ are superfields.

The Taylor series expansion for the superfields
${I\!\!N}(t,\eta,\bar\eta)$ and ${I\!\!R}(t,\eta,\bar\eta)$ are
the following
\begin{eqnarray}
{I\!\! N}(t,\eta,\bar\eta)&=& N(t) + i\eta\bar\psi^{\prime}(t) +
i\bar
\eta\psi^{\prime}(t) + V^{\prime}(t)\eta\bar\eta,\label{8}\\
{I\!\! R}(t,\eta,\bar\eta) &=& R(t) +
i\eta\bar\lambda^{\prime}(t)+
i\bar\eta\lambda^{\prime}(t) + B^{\prime}(t) \eta\bar\eta.\label{9}%
\end{eqnarray}
In the expressions (\ref{8}) and (\ref{9}) we have introduced the
redefinitions $\psi^{\prime}(t) = N^{1/2}\psi(t)$, $V^{\prime} =
N(t)V(t) + \bar\psi(t) \psi(t)$, $\lambda^{\prime} = \frac{\tilde
G^{1/2}N^{1/2}}{cR^{1/2}} \lambda$ and $B^{\prime} = \frac{\tilde
G^{1/2}}{c}NB + \frac{\tilde G^{1/2}}{2c R^{1/2}}(\bar\psi \lambda
- \psi \bar\lambda)$. The components of the superfield
${I\!\!N}(t, \eta, \bar\eta)$ are gauge fields of the
one-dimensional $n=2$ extended supergravity. $N(t)$ is the
einbein, $\psi(t), \bar\psi(t)$ are the complex gravitino fields,
and $V(t)$ is the $U(1)$ gauge field. The component $B(t)$ in
(\ref{9}) is an auxiliary degree of freedom (non-dynamical
variable), and $\lambda, \bar\lambda$ are the fermion partners of
the scale factor $R(t)$. After the integration over the Grassmann
coordinates $\theta, \bar\theta$ we can rewrite the action
(\ref{6}) in its component form
\begin{eqnarray}
S_{susy} &=& \int \left\{- \frac{c^2 R (DR)^2}{2N \tilde G} +
\frac{i}{2}(\bar\lambda D\lambda - D\bar\lambda \lambda) -
\frac{NR}{2}B^2 - \frac{N \tilde G^{1/2} B}{2cR}\bar\lambda
\lambda + \right. \nonumber\\
&& \left. + \frac{c^2 \sqrt{k} RN}{\tilde G^{1/2}} B + \frac{c^2
\sqrt{k} R^{1/2}}{2\tilde G^{1/2}}(\bar\psi \lambda - \psi
\bar\lambda) + \frac{cN \sqrt{k}}{R}\bar\lambda \lambda + \right.
\label{10}\\
&& \left. + \frac{c^2 \Lambda^{1/2}}{\sqrt{3} \tilde G^{1/2}}
NR^2B + \frac{c^2 \Lambda^{1/2} R^{3/2}}{2\sqrt{3} \tilde G^{1/2}}
(\bar\psi \lambda - \psi \bar\lambda) + \frac{2 c \Lambda^{1/2}
N}{\sqrt{3}} \bar\lambda \lambda - \right. \nonumber\\ && \left. -
\sqrt{2}cM^{1/2}_{\gamma}N R^{\frac{1 - 3\gamma}{2}} B -
\frac{\sqrt{2}}{2} c
M_{\gamma}^{1/2}R^{-\frac{3\gamma}{2}}(\bar\psi \lambda - \psi
\bar\lambda) - \right. \nonumber\\
&& \left. - \sqrt{2}(1 - 3\gamma)\tilde G^{1/2}M_{\gamma}^{1/2}N
R^{\frac{-3 - 3\gamma}{2}}\bar\lambda \lambda \right
\}dt.\nonumber
\end{eqnarray}
So, the lagrangian for the auxiliary field has the form
\begin{eqnarray}
L_B &=& -\frac{N R}{2} B^2 - \frac{N \tilde G^{1/2} B}{2 cR}
\bar\lambda \lambda  + \frac{c^2 \sqrt{k} RN}{\tilde G^{1/2}}B +
\frac{c^2 \Lambda^{1/2} N R^2 }{\sqrt{3} \tilde G^{1/2}}B -
\nonumber\\
&& - \sqrt{2} c M_{\gamma}^{1/2} NR^{\frac{1 - 3\gamma}{2}} B.
\label{11}
\end{eqnarray}
From the expression (\ref{11}) we can obtain the equation for the
auxiliary field varying the Lagrangian with respect to $B$
\begin{equation}
B = \frac{c^2 \sqrt{k}}{ \tilde G^{1/2}} -\frac{ \tilde
G^{1/2}}{2cR^2}\bar\lambda \lambda + \frac{c^2 \Lambda^{1/2} R}{
\sqrt{3} \tilde G^{1/2}} - \sqrt{2} c M_{\gamma}^{1/2}
R^{\frac{-3\gamma - 1}{2}}. \label{12}
\end{equation}
Then, putting the expression (\ref{12}) in (\ref{10}) we have the
following supersymmetric action
\begin{eqnarray}
S_{susy} &=& \int \left\{- \frac{c^2 R (DR)^2}{2N \tilde G} +
\frac{c^4 NkR}{2\tilde G} + \frac{c^4 N \Lambda R^3}{6 \tilde G} +
Nc^2M_\gamma R^{-3\gamma} + \right.
\nonumber\\
&& \left. + \frac{c^4 \sqrt{k} \Lambda^{1/2} R^2}{\sqrt{3} \tilde
G} -  \frac{\sqrt{2k}c^3}{\tilde G^{1/2}}
M_{\gamma}^{1/2}R^{\frac{1 - 3\gamma}{2}} -
\frac{\sqrt{2}c^3\Lambda^{1/2} M_\gamma^{1/2}}{\sqrt{3} \tilde
G^{1/2}} R^{\frac{3 - 3 \gamma}{2}} + \right.
\nonumber\\
&& \left. + \frac{i}{2}(\bar\lambda D\lambda - D\bar\lambda
\lambda)  + \frac{cN \sqrt{k}}{2R} \bar\lambda \lambda +
\frac{\sqrt{3}}{2}c\Lambda^{1/2}N \bar\lambda \lambda + \right.
\label{13}\\
&& \left. + \frac{(-1 + 6\gamma)}{\sqrt{2}} N \tilde G^{1/2}
M_\gamma^{1/2}R^{\frac{-3 - 3 \gamma}{2}}\bar\lambda \lambda +
\frac{c^2 \sqrt{k} R^{1/2}}{2\tilde G^{1/2}}(\bar\psi \lambda -
\psi \bar\lambda)\right.
\nonumber\\
&& \left. + \frac{c^2 \Lambda^{1/2}}{2\sqrt{3}\tilde G^{1/2}}
R^{3/2}(\bar\psi \lambda - \psi \bar\lambda) - \frac{\sqrt{2}}{2}
cM_\gamma^{1/2}R^{-\frac{3\gamma}{2}}(\bar\psi \lambda - \psi
\bar\lambda) \right\}dt,  \nonumber
\end{eqnarray}
where $DR = {\dot R} - \frac{i\tilde G^{1/2}}{2cR^{1/2}}(\psi
\bar\lambda + \bar\psi \lambda)$ and $D\lambda = {\dot \lambda} -
\frac{1}{2} V\lambda$, $D{\bar\lambda} = {\dot {\bar\lambda}} +
\frac{1}{2} V\bar\lambda$.
\begin{center}
{\bf Supersymmetric Quantum Model}
\end{center}
In this section we will proceed with the quantization analysis of
the system. The classical canonical Hamiltonian is calculated in
the usual way for the systems with constraints. It has the form
\begin{equation}
H_c = NH + \frac{1}{2} \bar\psi S - \frac{1}{2}\psi {\bar S} +
\frac{1}{2}VF, \label{14}
\end{equation}
where $H$ is the Hamiltonian of the system, $S$ and ${\bar S}$ are
the supercharges and $F$ is the $U(1)$ rotation generator. The
form of the canonical Hamiltonian (\ref{14}) explains the fact
that $N, \psi, \bar\psi$ and $V$ are Lagrangian multipliers which
only enforce the first-class constraints $H = 0, S = 0, {\bar S }
= 0$ and $F = 0$, which express the invariance under the conformal
$n = 2$ supersymmetric transformations. The first-class
constraints may be obtained from the action (\ref{13}) varying
$N(t),\psi(t)$,$\bar\psi(t)$ and $V(t)$, respectively. The
first-class constraints are
\begin{eqnarray}
H &=& -\frac{\tilde G}{2c^2 R} \pi^2_R - \frac{c^4 k R}{2\tilde G}
- \frac{c^4 \Lambda R^3}{6 \tilde G} - M_{\gamma}c^2R^{-3\gamma} +
\frac{\sqrt{2}c^3\Lambda^{1/2} M_\gamma^{1/2}}{\sqrt{3} \tilde
G^{1/2}} R^{\frac{3 - 3
\gamma}{2}} - \nonumber\\
&-& \frac{c^4 \sqrt{k} \Lambda^{1/2} R^2}{\sqrt{3} \tilde G} +
\frac{\sqrt{2k}c^3}{\tilde G^{1/2}} M_{\gamma}^{1/2}R^{\frac{1 -
3\gamma}{2}} - \frac{c \sqrt{k}}{2R} \bar\lambda \lambda -
\frac{\sqrt{3}}{2}c\Lambda^{1/2} \bar\lambda \lambda -
\nonumber\\
&-& \frac{(6\gamma - 1)}{\sqrt{2}}\tilde G^{1/2}
M_\gamma^{1/2}R^{\frac{-3 - 3 \gamma}{2}}\bar\lambda \lambda
,\label{15}\\
S&=& \Big(\frac{i\tilde G^{1/2}}{c R^{1/2}}\pi_R - \frac{c^2
\sqrt{k} R^{1/2}}{\tilde G^{1/2}} - \frac{c^2 \Lambda^{1/2}
R^{3/2}}{\sqrt{3}\tilde G^{1/2}} + \sqrt{2} c M_{\gamma}^{1/2}
R^{-\frac{3\gamma}{2}}\Big)
\lambda, \label{16} \\
{\bar S}&=& \Big(-\frac{i\tilde G^{1/2}}{c R^{1/2}}\pi_R -
\frac{c^2 \sqrt{k} R^{1/2}}{\tilde G^{1/2}} - \frac{c^2
\Lambda^{1/2} R^{3/2}}{\sqrt{3}\tilde G^{1/2}} + \sqrt{2} c
M_{\gamma}^{1/2} R^{-\frac{3\gamma}{2}}\Big)\bar\lambda, \label{17}\\
F&=& - \bar\lambda \lambda, \label{18}
\end{eqnarray}
where $\pi_R = - \frac{c^2 R}{{\tilde G}N} {\dot R} +
\frac{icR^{1/2}}{2N \tilde G^{1/2}}(\bar\psi \lambda + \psi
\bar\lambda)$ is the canonical momentum associated to $R$. The
canonical Dirac brackets are defined as
\begin{equation}
\lbrace R, \pi_R \rbrace = 1,\quad \lbrace \lambda, \bar\lambda
\rbrace = i. \label{19}
\end{equation}
With respect to these brackets the super-algebra for the
generators $H, S, {\bar S}$ and $F$ becomes
\begin{equation}
\lbrace S, {\bar S}\rbrace = - 2iH,\quad \lbrace S, H \rbrace =
\lbrace {\bar S}, H \rbrace = 0, \quad \lbrace F, S\rbrace =
iS,\quad \lbrace F, {\bar S}\rbrace = i{\bar S}. \label{20}
\end{equation}
In a quantum theory the brackets (\ref{19}) must be replaced by
anticommutators and commutators, they can be considered as
generators of the Clifford algebra. We have
\begin{eqnarray}
\lbrace \lambda, \bar\lambda \rbrace &=& -\hbar, \qquad [R,
\pi_R]= i\hbar \, \qquad {\rm with} \quad \
\pi_R=-i\hbar \frac{\partial}{\partial R} \label{21} \\
\bar \lambda &=&  \xi^{-1} \lambda^{\dag} \xi =- \lambda^{\dag},
\qquad \lbrace \lambda, \lambda^{\dag} \rbrace =  \hbar, \qquad
\lambda^{\dag} \xi= \xi \lambda^{\dag} \quad {\rm and} \quad
\xi^{\dag}=\xi. \nonumber
\end{eqnarray}
Then, for the operator ${\bar S}$ the following equation is
satisfied
\begin{equation}
\bar S=\xi^{-1} S^{\dag} \xi. \label{22}
\end{equation}
Therefore, the anticommutator of supercharges $S$ and their
conjugated operator ${\bar S}$ under our defined conjugation has
the form
\begin{equation}
\overline {\left\{ S, \bar S \right\}}= \xi^{-1} \left\{ S, \bar S
\right\} \xi = \left\{ S, \bar S \right\}, \label{23}
\end{equation}
and the Hamiltonian operator is self-conjugated under the
operation ${\bar H} = \xi^{-1} H^{\dagger} \xi$. We can choose the
matrix representation for the fermionic para\-meters $\lambda,
\bar\lambda$ and $ \xi$ as
\begin{equation}
\lambda = \sqrt{\hbar} \sigma_-, \qquad \bar \lambda = -
\sqrt{\hbar} \sigma_+, \qquad \xi= \sigma_3, \label{24}
\end{equation}
with $\sigma_\pm = \frac{1}{2}(\sigma_1 \pm i\sigma_2)$, where
$\sigma_1$, $\sigma_2$, $\sigma_3$ are the Pauli matrices.
\\

In the quantum level we must consider the nature of the Grassmann
variables $\lambda$ and $\bar\lambda$, with respect to these we
perform the antisymmetrization, then we can write the bilinear
combination in the form of the commutators, $\bar\lambda, \lambda
\to \frac{1}{2}[\bar\lambda, \lambda]$, and this leads to the
following quantum Hamiltonian $H$.
\begin{eqnarray}
H_{quantum} &=& -\frac{\tilde G}{2c^2} R^{-1/2} \pi_R R^{-1/2}
\pi_R - \frac{c^4 k R}{2\tilde G} - \frac{c^4 \Lambda R^3}{6
\tilde G} - M_{\gamma}c^2R^{-3\gamma} \nonumber\\
&+& \frac{\sqrt{2}c^3\Lambda^{1/2} M_\gamma^{1/2}}{\sqrt{3} \tilde
G^{1/2}} R^{\frac{3 - 3 \gamma}{2}} - \frac{c^4 \sqrt{k}
\Lambda^{1/2} R^2}{\sqrt{3} \tilde G} + \nonumber\\
&+& \frac{\sqrt{2k}c^3}{\tilde G^{1/2}} M_{\gamma}^{1/2}R^{\frac{1
- 3\gamma}{2}} - \frac{c \sqrt{k}}{4R}[\bar\lambda, \lambda] -
\frac{\sqrt{3}}{4}c\Lambda^{1/2} [\bar\lambda, \lambda] -
\nonumber\\
&-& \frac{(6\gamma - 1)}{2\sqrt{2}}\tilde G^{1/2}
M_\gamma^{1/2}R^{\frac{-3 - 3 \gamma}{2}}[\bar\lambda, \lambda].
\label{25}
\end{eqnarray}
The supercharges $S$, $\bar S$ and the fermion number $F$ have the
following structures:
\begin{equation}
S = A\lambda, \qquad\qquad S^{\dag} = A^{\dag} \lambda^{\dag}
\label{26}
\end{equation}
where
\begin{equation}
A = \frac{i\tilde G^{1/2}}{c} R^{-1/2} \pi_R - \frac{c^2
\sqrt{k}}{\tilde G^{1/2}} R^{1/2} - \frac{c^2 \Lambda^{1/2}
R^{3/2}}{\sqrt{3}\tilde G^{1/2}} + \sqrt{2}
cM_\gamma^{1/2}R^{-\frac{3\gamma}{2}},\label{27}
\end{equation}
and
\begin{equation}
F = -\frac{1}{2}[\bar\lambda, \lambda]. \label{28}
\end{equation}
An ambiguity exist in the factor ordering of these operators, such
ambiguities always arise, when the operator expression contains
the product of non-commuting operator $R$ and $\pi_R$, as in our
case. It is then necessary to find some criteria to know which
factor ordering should be selected. The inner product is
calculated performing the integration with the measure $R^{1/2}
dR$. With this measure the conjugate momentum $\pi_R$ is
non-Hermitian with $\pi^{\dagger}_R = R^{-1/2} \pi_R R^{1/2}$.
However, the combination $(R^{-1/2} \pi_R)^{\dagger} =
\pi_R^{\dagger} R^{-1/2} = R^{-1/2} \pi_R$ is a Hermitian one, and
$(R^{-1/2} \pi_R R^{1/2} \pi_R)^{\dagger} = R^{-1/2}\pi_R
R^{1/2}\pi_R$ is Hermitian too. This choice in our supersymmetric
quantum approach $n = 2$ eliminates the factor ordering ambiguity
by fixing the ordering parameter $p = \frac{1}{2}$.
\begin{center}
{\bf Superquantum Solutions}
\end{center}
In the quantum theory, the first-class constraints $H = 0, S = 0,
{\bar S} = 0$ and $F = 0$ become conditions on the wave function
$\Psi(R)$. Furthermore, any physical state must be satisfied the
quantum constraints
\begin{equation}
H \Psi(R) = 0, \quad S\Psi(R) = 0,\quad {\bar S}\Psi(R) = 0, \quad
F\Psi(R) = 0, \label{29}
\end{equation}
where the first equation is the Wheeler-DeWitt equation for the
minisuperspace model. The eigenstates of the Hamiltonian
(\ref{25}) have two components in the matrix representation
(\ref{24})
\begin{equation}
\Psi = \pmatrix{\Psi_1\cr \Psi_2 \cr}. \label{30}
\end{equation}
However, the supersymmetric physical states are obtained applying
the supercharges operators $S\Psi = 0, {\bar S}\Psi = 0$. With the
conformal algebra given by (\ref{20}), these are rewritten in the
following form
\begin{equation}
(\lambda {\bar S} - \bar\lambda S)\Psi = 0.\label{31}
\end{equation}
Using the matrix representation for $\lambda$ and $\bar\lambda$ we
obtain the following differential equations for $\Psi_1(R)$ and
$\Psi_2(R)$ components
\begin{eqnarray}
\Big(\frac{\hbar \tilde G^{1/2}}{c} R^{-1/2} \frac{\partial
}{\partial R} - \frac{c^2 \sqrt{k} R^{1/2}}{\tilde G^{1/2}} -
\frac{c^2 \Lambda^{1/2} R^{3/2}}{\sqrt{3} \tilde G^{1/2}} +
\sqrt{2} cM_\gamma^{1/2}R^{-\frac{3\gamma}{2}}
\Big) \Psi_1(R) = 0.\label{32}\\
\Big(\frac{\hbar \tilde G^{1/2}}{c} R^{-1/2} \frac{\partial
}{\partial R} + \frac{c^2 \sqrt{k} R^{1/2}}{\tilde G^{1/2}} +
\frac{c^2 \Lambda^{1/2} R^{3/2}}{\sqrt{3} \tilde G^{1/2}} -
\sqrt{2} cM_\gamma^{1/2}R^{-\frac{3\gamma}{2}} \Big) \Psi_2(R) =
0.\label{33}
\end{eqnarray}
Solving these equation, we have the following wave functions
solutions
\begin{eqnarray}
\Psi_1(R) = C \exp{\Big[ \frac{\sqrt{k} c^{3}R^{2}}{2\hbar \tilde
G} + \frac{c^3 \Lambda^{1/2}}{3 \sqrt{3}\hbar \tilde G}R^{3} -
\frac{2\sqrt{2}c^2M_{\gamma}^{1/2}}{(3 - 3\gamma)\hbar \tilde
G^{1/2}}R^{\frac{3 - 3 \gamma}{2}}\Big]},\\ \label{34} \Psi_2(R) =
\tilde C \exp{\Big[ -\frac{\sqrt{k} c^{3}R^{2}}{2\hbar \tilde G} -
\frac{c^3 \Lambda^{1/2}}{3 \sqrt{3}\hbar \tilde G}R^{3} +
\frac{2\sqrt{2}c^2M_{\gamma}^{1/2}}{(3 - 3\gamma)\hbar \tilde
G^{1/2}}R^{\frac{3 - 3 \gamma}{2}}\Big]}.
\end{eqnarray}
In the case of the flat universe $(k = 0)$ and for the dust-like
matter $(\gamma = 0)$ we have the following solutions (using the
relation $M_{\gamma = 0} = \frac{1}{2}R^3\rho_{\gamma = 0})$
\begin{eqnarray}
\Psi_1(R) = C_1 \exp{\Big[ \frac{1}{\sqrt{6\pi}}\Big(
\frac{\rho_\Lambda}{\rho_{pl}}\Big)^{1/2} \Big(
\frac{R}{l_{pl}}\Big)^3 - \frac{\sqrt{2}}{\sqrt{6\pi}} \Big(
\frac{\rho_{\gamma = 0}}{\rho_{pl}}\Big)^{1/2} \Big(
\frac{R}{l_{pl}}}\Big)^3\Big], \label{35}
\end{eqnarray}
\begin{eqnarray}
\Psi_2(R) = C_2 \exp{\Big[ -\frac{1}{\sqrt{6\pi}}\Big(
\frac{\rho_\Lambda}{\rho_{pl}}\Big)^{1/2} \Big(
\frac{R}{l_{pl}}\Big)^3 + \frac{\sqrt{2}}{\sqrt{6\pi}} \Big(
\frac{\rho_{\gamma = 0}}{\rho_{pl}}\Big)^{1/2} \Big(
\frac{R}{l_{pl}}}\Big)^3\Big], \label{36}
\end{eqnarray}
where $\rho_{pl} = \frac{c^5}{\hbar G^2}$ is the Planck density
and $l_{pl} = \Big(\frac{\hbar G}{c^3}\Big)^{1/2}$ is the Planck
length.
\\

We can see, that the function $\Psi_1$ in (\ref{35}) has good
behavior when $R \to \infty$ under the condition $\rho_\Lambda < 2
\rho_{\gamma = 0}$, while $\Psi_2$ does not. On the other hand,
the wave function $\Psi_2$ in (\ref{36}) has good behavior when $R
\to \infty$ under the condition $\rho_\Lambda > 2 \rho_{\gamma =
0}$, because the principal contribution comes from the first term
of the exponent, while $\Psi_1$ does not have good behavior.
However, only the scalar product for the second wave function
$\Psi_2$ is normalizable in the measure $R^{1/2}dR$ under the
condition $\rho_\Lambda > 2 \rho_{\gamma = 0}$. This condition
does not contradict the astrophysical observation at
$\rho_{\Lambda} \approx (2 - 3)\rho_{M}$, due to the fact that the
dust matter introduces the main contribution to the total energy
density of matter $\rho_M$.
\\

On the other hand, according to recent astrophysical data, our
universe is dominated by a mysterious form of the dark energy
\cite{4}, which counts to about $70$ per cent of the total energy
density. As a result, the universe expansion is accelerating
\cite{5,6}. Vacuum energy density $\rho_\Lambda = \frac{c^2
\Lambda}{8\pi G}$ is a concrete example of the dark energy.
\begin{center}
{\bf Conclusion}
\end{center}
The recent cosmological data give us the following range for the
dark energy state parameter $\gamma = - 0.96^{+0.08}_{-0.09}$.
However, in the literature we can find different theoretical
models for the dark energy with state parameter $\gamma > - 1$ and
$\gamma < - 1$, see reviews \cite{7,8} and the articles
\cite{9,10}. In the present work we have discussed the case for
$\gamma = 0$ corresponding to the FRW universe with barotropic
perfect fluid as matter field. In the case of the flat universe $(
k = 0)$ and the dust-like matter $\gamma = 0$ we have obtained two
wave functions. However, only the second wave function is
normalizable under the condition $\rho_\Lambda > 2\rho_{\gamma =
0}$, which leads to the cosmological value $\Lambda
> \frac{16 \pi G}{c^2}\rho_{\gamma = 0}.$
\\


\begin{thebibliography}{99}

\bibitem{1} O. Obreg\'on, J.J. Rosales and V.I. Tkach, Phys. Rev.
{\bf D 53}, R1750 (1996); V.I. Tkach, J.J. Rosales and O.
Obreg\'on, Class. Quantum Grav. {\bf 13}, 2349 (1996).

\bibitem{2} C. Ortiz, J.J. Rosales, J. Socorro, J. Torres and V.I.
Tkach. Phys. Lett. {\bf A 340}, 51-58, (2005); O. Obreg\'on, J.J.
Rosales, J. Socorro and V.I. Tkach, Class. Quantum Gra\-vity, {\bf
16}, 2861 (1999); V.I. Tkach, J.J. Rosales and J. Socorro, Class.
Quantum Gravity, {\bf 16}, 797 (1999); M. Ryan Jr, Hamiltonian
Cosmology, Springer Verlag (1972); J. Socorro, M.A. Reyes and F.A.
Gelbert, Phys. Lett. {\bf A 313}, 338 (2003).

\bibitem{3} V.I. Tkach, arXiv:0808.3429; C. Beck, arXiv:0810.0752.

\bibitem{4} T. Padmanabhan, Phys. Rept. {\bf 380}, 235 (2003).

\bibitem{5} S. Perlmutter, et al; Astrophysics J., {\bf 517}, 565
(1999).

\bibitem{6} A.G. Riess, et al; Astrophysics J., {\bf 607}, 665
(2004).

\bibitem{7} S. Nojiri, S.D. Odintsov, Int. J. Geom. Math. Mod.
Phys. {\bf 4}, 115 (2007) and references therein.

\bibitem{8} E.J. Copeland, M. Sami, S. Tsujikawa, Int. J. Mod.
Phys. {\bf D} 15, 1753 (2006) and references therein.

\bibitem{9} S. Nojiri, S.D. Odintsov, arXiv: 0801.4843; arXiv:
0807.0685; arxiv: 0810.1557.

\bibitem{10} K. Bamba, C. Geng, S. Nojiri, S.D, Odinsov,
arxiv:0810.4296
\end{thebibliography}
\end{document}